\RequirePackage{fix-cm}
\documentclass{svjour3}                     
\smartqed  
\usepackage{graphicx}
\usepackage{amsmath, latexsym, amsfonts, amssymb, subfigure, cite}
\newcommand{\ra}{\rangle}
\newcommand{\la}{\langle}

\newcommand{\re}{\textup{Re}}

\newcommand{\wh}{\widehat}

\newcommand{\eAp}{e^{-\frac{i}{\hbar}\gamma\wh{A}\wh{p}}}
\newcommand{\0}{\tilde{0}}
\newcommand{\1}{\tilde{1}}
\newcommand{\tu}{\textup}
\begin{document}

\title{A note concerning the modular valued von Neumann interaction operator
}


\author{A.D. Parks         \and
        S.E. Spence        \and 
        J.M. Farinholt  
}


\institute{A.D. Parks \at
              Electromagnetic \& Sensor Systems Department \\
              Naval Surface Warfare Center, Dahlgren Division\\
              Tel.: +1-540-653-0582\\
              \email{allen.parks@navy.mil}           
           \and
           S.E. Spence \at
              Electromagnetic \& Sensor Systems Department \\
              Naval Surface Warfare Center, Dahlgren Division\\
              Tel.: +1-540-653-1577\\
              \email{scott.e.spence@navy.mil}           
           \and
           J.M. Farinholt \at
              Strategic \& Computing Systems Department \\
              Naval Surface Warfare Center, Dahlgren Division\\
              Tel.: +1-540-653-9576\\
              \email{jacob.farinholt@navy.mil}        
}

\date{Received: date / Accepted: date}

\maketitle

\begin{abstract}
The modular valued operator $\widehat{V}_m$ of the von Neumann interaction operator for a projector is defined. The properties of $\widehat{V}_m$ are discussed and contrasted with those of the standard modular value of a projector. The associated notion of a faux qubit is introduced and its possible utility in quantum computation is noted. An experimental implementation of $\widehat{V}_m$ is also highlighted.
\end{abstract}

\medskip
The notion of the quantum modular value of a projection operator was introduced in 2010 by Kedem and Vaidman \cite{KedemVaidman10}. Because of the intrinsic relationship of such modular values to weak values, quantum foundations, and quantum computing, recent studies have been devoted to illuminating many of their properties \cite{CorEtAl16, HoImoto16a, HoImoto16b, CorCuad17, HoImoto17, BroCoh17}. This short note defines the ``modular valued operator'' $\widehat{V}_m$ of the von Neumann interaction operator when the associated observable $\widehat{A}$ to be measured is a projector. The properties of $\widehat{V}_m$ are discussed and contrasted with those of the modular value of $\widehat{A}$ and it is shown that - for arbitrary interaction strength - the action of $\widehat{V}_m$ upon a continuous pointer state produces a new state which mimics the behavior of the well-known two slit experiment. This new state can also serve as a “faux” (fake) qubit. The potential utility of faux qubits in quantum computing is noted and a recent experimental implementation of $\widehat{V}_m$ is briefly described.

The \emph{von Neumann interaction operator} $\wh{V}$ describes the coupling between a quantum system and a measurement pointer according to
\begin{equation}\label{Eq:vN_Op}
\wh{V} \equiv e^{-\frac{i}{\hbar}\gamma \wh{A}\wh{p}},
\end{equation}
where $\gamma = \int \gamma(t)dt$ is the coupling strength of  the interaction, $\wh{A}$ is the system observable to be measured, and $\wh{p}$ is the pointer's momentum operator. For typical measurements the pointer variable $\wh{p}$ and the initial pointer state $|\varphi\rangle$ employed are continuous so that if the system is pre-selected to be in state $|\psi_i\rangle$, the pointer state after the measurement is $\wh{V}|\psi_i\ra|\varphi\ra$. If state $|\psi_f\ra$ is post-selected after the interaction, the resulting pointer state of the pre- and post-selected (PPS) system is $|\Phi\ra = \la\psi_f |\wh{V}|\psi_i\ra |\varphi\ra$. In this context, the weak value of $\wh{A}$ occurs when the coupling strength is sufficiently small so that $\la\psi_f |\wh{V}|\psi_i\ra \approx e^{-\frac{i}{\hbar}\gamma A_w \wh{p}}$ and $|\Phi\ra \approx e^{-\frac{i}{\hbar}\gamma A_w \wh{p}}|\varphi\ra$ - in which case the pointer is shifted from its initial position by the amount $\gamma \textup{Re}A_w$. Here the complex number
\begin{equation}
A_w = \frac{\la\psi_f|\wh{A}|\psi_i\ra}{\la\psi_f|\psi_i\ra}
\end{equation}
is the weak value of $\wh{A}$.

The \emph{modular value of} $\wh{A}$ arises when the initial pointer state is a normalized qubit $|\varphi\ra = \alpha|0\ra + \beta|1\ra$ and $\wh{p}$ in Eq. \eqref{Eq:vN_Op} is replaced by the projector $|1\ra\la 1|$. After a measurement with arbitrary coupling strength is performed upon a PPS system, the resulting normalized state is the qubit
\begin{align}
|\Phi\ra &= \frac{1}{N}\la\psi_f|\wh{V}|\psi_i\ra|\varphi\ra \\
         &= \frac{1}{N}\left[\alpha \la\psi_f|e^0 |\psi_i\ra|0\ra + \beta \la\psi_f|e^{-\frac{i}{\hbar}\gamma \wh{A}}|\psi_i\ra |1\ra \right],
\end{align}
or
\begin{equation}\label{Eq:MV_PointerShift}
|\Phi\ra = \frac{1}{N} \la\psi_f|\psi_i\ra \left[ \alpha|0\ra + \beta (\wh{A})_m |1\ra \right],
\end{equation}
where $N$ is the normalization factor and the complex number
\begin{equation}
(\wh{A})_m \equiv \frac{\la\psi_f|e^{-\frac{i}{\hbar}\gamma\wh{A}}|\psi_i\ra}{\la\psi_f|\psi_i\ra}
\end{equation}
is the modular value of $\wh{A}$. Although $(\wh{A})_m$ is related to $A_w$ via the approximate expression $(\wh{A})_m \approx 1-\frac{i}{\hbar}\gamma A_w \approx e^{-\frac{i}{\hbar}\gamma A_w}$ when $\gamma$ is sufficiently small, the formal general relationship between them is given exactly by the derivative \cite{HoImoto16b}
\begin{equation}
A_w = i\hbar \frac{\partial(\wh{A})_m}{\partial\gamma}\Biggr\rvert_{\gamma=0}.
\end{equation}

Now extend the notion of a quantum modular value to the \emph{modular valued operator} $\wh{V}_m$ of $\wh{V}$ defined by
\begin{equation}
\wh{V}_m \equiv \frac{\la\psi_f|e^{-\frac{i}{\hbar}\gamma\wh{A}\wh{p}}|\psi_i\ra}{\la\psi_f|\psi_i\ra},
\end{equation}
where $\wh{A}$ is understood to be a projector. Since $\eAp = \wh{1} - \wh{A} + \wh{A}\wh{S}$ when $\wh{A}$ is a projector and $\wh{p}$ (and $|\varphi\ra$) are continuous \cite{ParksEtAl14}, then - for arbitrarily large $\gamma$ - the operator $\wh{V}_m$ can be exactly expressed in terms of $A_w$ as
\begin{equation}
\wh{V}_m = (1-A_w)\wh{1} + A_w\wh{S},
\end{equation}
where $\wh{S} = e^{-\frac{i}{\hbar}\gamma\wh{p}}$ is the pointer position translation operator defined by the action $\la q|\wh{S}|\varphi\ra \equiv \varphi(q-\gamma)$, $\wh{q}$ is the pointer's position operator that is conjugate to $\wh{p}$, and - for simplicity - it is assumed that $\la q|\varphi\ra \equiv \varphi(q)$ is real valued and symmetric about $\la \varphi |\wh{q}|\varphi\ra = 0$ (here $\wh{1}$ is the identity operator).

The \emph{exact} expression for the normalized pointer state $|\Phi\ra$ that results from the action of $\wh{V}_m$ upon $|\varphi\ra$ is (Eq. 2.3 in \cite{ParksEtAl14})
\begin{equation}\label{Eq:PointerExact}
|\Phi\ra = \frac{e^{i\chi}}{M}\left[ (1-A_w)|\varphi\ra + A_w\wh{S}|\varphi\ra \right],
\end{equation}
where $\chi$ is the Pancharatnam phase defined by $e^{i\chi} = \la\psi_f|\psi_i\ra |\la\psi_f|\psi_i\ra|^{-1}$ and
\begin{equation*}
M = \left[ 1-2\re A_w + 2|A_w|^2 + A_w(1-A_w^*)\la\varphi|\wh{S}|\varphi\ra + A_w^*(1-A_w)\la\varphi|\wh{S}^\dag |\varphi\ra \right]^{1/2}.
\end{equation*}
Thus, similar to the qubit resulting from an arbitrarily strong PPS measurement of a projector $\wh{A}$ which has the modular value of $\wh{A}$ encoded within it, an arbitrarily strong action of $\wh{V}_m$ upon a continuous pointer state produces a new pointer state with $A_w$ encoded within it.

The exact spatial distribution profile for this new pointer state is
\begin{equation}\label{Eq:SpatialDistro}
\Phi^2(q) = \left(\frac{1}{M^2}\right)\{|1-A_w|^2 \varphi(q)^2 + |A_w|^2 \varphi(q-\gamma)^2 + 2\re[A_w(1-A_w^*)] \varphi(q)\varphi(q-\gamma)\}.
\end{equation}
Observe that if $A_w \in \{0, 1\}$, then $M=1$, the interference term $2\re[A_w(1-A_w^*)]\varphi(q)\varphi(q-\gamma)$ in Eq. \eqref{Eq:SpatialDistro} vanishes, and the distribution profiles are given exactly by
\begin{equation}\label{Eq:SpDistro_deginerate}
\Phi(q)^2 = 
\begin{cases}
\varphi(q)^2, & A_w = 0\\
\varphi(q-\gamma)^2, & A_w = 1
\end{cases},
\end{equation}
whereas if $A_w \not\in \{0, 1\}$, this interference term does not vanish and the profile is given by Eq. \eqref{Eq:SpatialDistro}. It is interesting that this change of $\Phi(q)^2$ with the value of $A_w$ is reminiscent of the changes that occur in the two-slit experiment when parallel slits located along a line on a screen at positions $q$ and $q-\gamma$ are blocked ($A_w \in \{0,1\}$) and there is no interference and unblocked ($A_w \not\in \{0,1\}$) and there is interference. As a side note, Eq. \eqref{Eq:SpDistro_deginerate} is a formal statement that the action of $\wh{V}_m$ upon $|\varphi\ra$ produces a state that exhibits \emph{weak value persistence} (recall that weak value persistence refers to the fact that the measured weak values 0 and 1 for a projector persist for large coupling strengths outside the weak measurement regime where the measurements are no longer weak measurements) \cite{ParksEtAl14, ParksSpence16}.

The similarity between the form of Eq. \eqref{Eq:PointerExact} and that of Eq. \eqref{Eq:MV_PointerShift} suggest that Eq. \eqref{Eq:PointerExact} can be rewritten as the \emph{faux qubit}
\begin{equation}
|\Phi\ra = \frac{e^{i\chi}}{M}\left[(1-A_w)|\tilde{0}\ra + A_w|\tilde{1}\ra\right],
\end{equation}
where the substitutions $|\varphi\ra = |\0\ra$ and $\wh{S}|\varphi\ra = |\1\ra$ have been made in Eq. \eqref{Eq:PointerExact}. The adjective ``faux'' is used here because, although $\la\0|\0\ra = \la\varphi|\varphi\ra = 1 = \la\varphi|\varphi\ra = \la\varphi|\wh{S}^\dag \wh{S}|\varphi\ra = \la\1|\1\ra$, $|\0\ra$ and $|\1\ra$ are orthogonal only in the limit of infinitely large coupling strengths, i.e.
\begin{equation}
\lim_{\gamma\rightarrow\infty}\la\0|\1\ra = \lim_{\gamma\rightarrow\infty}\la\varphi|\wh{S}|\varphi\ra = 0 = \lim_{\gamma\rightarrow\infty}\la \varphi |\wh{S}^\dag|\varphi\ra = \lim_{\gamma\rightarrow\infty}\la\1|\0\ra.
\end{equation}

This - along with the distribution profile of Eq. \eqref{Eq:SpatialDistro} - implies that ensemble measurement can be used as an approach to ``reading'' a faux qubit when $A_w$ is real valued. When the distribution width of $\varphi(q)^2$ is sufficiently narrow and the coupling strength is sufficiently large, the interference term in Eq. \eqref{Eq:SpatialDistro} can be made quite small. In this case, Eq. \eqref{Eq:SpatialDistro} is given to good approximation by
\begin{equation}
\Phi^2(q) \approx \left(\frac{1}{M^2}\right)\{|1-A_w|^2\varphi(q)^2 + |A_w|^2\varphi(q-\gamma)^2\}.
\end{equation}
The distribution $\varphi(q)^2$ centered at $q=0$ corresponds to the distribution of the faux state $|\0\ra$ and the distribution $\varphi(q-\gamma)^2$ centered at (and separated from $q=0$ by) $q=\gamma$ corresponds to the distribution of the faux state $|\1\ra$. Provided that $A_w \not\in \{0,1\}$, the ratio of the peak value of $\varphi(q-\gamma)^2$ to that of $\varphi(q)^2$ is the ratio $|A_w|^2/|1-A_w|^2$ from which $A_w$ can be readily determined when it is real valued. In fact, in \cite{GKP01} a method to encode the above faux states into logical qubit states was given, so that is is possible to translate the (continuous variable) pointer state to a (logical) qubit framework. This suggests that $\wh{V}_m$ may have some utility in quantum computing.

Before closing, it is instructive to consider the twin Mach-Zehnder interferometer (MZI) which has been utilized to perform several recent experiments \cite{SpenceParks12, SpenceParks17} as an example of both an embodiment of $\wh{V}_m$ and a faux qubit reader when $A_w \in \{0, 1/2, 1\}$. Referring to Fig. 1 in \cite{SpenceParks17}, a classically intense monochromatic beam of laser light with Gaussian distribution (this defines the initial pointer state $|\varphi\ra$) is introduced into the MZI along path R1 (this defines$|\tu{R}1\ra$ as the pre-selected state). With a $\phi=0$ phase window, the apparatus is aligned so that the path L4-R5 is optically dark. A piezo-electrically driven computer controlled stage changes the location of M1 in the direction shown in Fig. 1 to produce a transverse spatial shift of the photon beam. The displacement of M1 is proportional to the coupling strength $\gamma$ and the reflection of the beam at M1 corresponds to the measurement of strength $\gamma$ of the projector $\wh{A}\equiv |\tu{L}2\ra\la \tu{L}2|$ (the apparatus is then re-aligned to keep the path L4-R5 optically dark). The beam traverses the optically bright path R4-L5 and emerges from BS3 along paths R6 and L6. A camera placed at R6 defines $|\tu{R}6\ra$ as the post-selected state, the image of the intensity distribution corresponds the distribution of the faux state $|\1\ra$, and the image's centroid to $\gamma A_w$ when $A_w = 1$. If the procedure is repeated with $\phi = \pi$, the distribution of the image corresponds to the distribution of the faux state $|\0\ra$ and the centroid of the image corresponds to $\gamma A_w$ when $A_w=0$ (note that if – instead - a camera is placed at L6 when $\phi=0$, then $|\tu{L}6\ra$ is the post-selected state, the image of the intensity distribution corresponds to the distribution of the faux state $|\0\ra$, and the associated centroid corresponds to $\gamma A_w$ when $A_w=0$). If $\phi=0$ and the dark path is blocked with a shutter, then the image of the intensity distribution at R6 corresponds to the distribution of the faux qubit $(|\0\ra + |\1\ra)$ and its centroid to $\gamma A_w$ when $A_w=1/2$. The observed pointer response with increasing coupling strength for $\phi\in\{0,\pi\}$ is provided in Fig. 3 in \cite{SpenceParks17} (the line labeled ``PSO...'' corresponds to the pointer response when L4-R5 is blocked \cite{SpenceParks12}). There the symmetric pointer response curves are a consequence of the decreasing overlap of beams along paths L3 and R3 at BS2 as the coupling strength increases. Note that faux qubits can only exist where these beams overlap since the overlap provides the retro-causal channel required to induce $A_w$. Additional detailed information concerning the interferometer, evaluation of weak values, and pointer response can be found in \cite{SpenceParks12, SpenceParks17, SPN12}.

\begin{acknowledgements}
This work was supported by Grants from the Naval Surface Warfare Center Dahlgren Division's In-house Laboratory Independent Research Program and the Naval Surface Warfare Center Dahlgren Division’s Naval Innovative Science and Engineering Program.
\end{acknowledgements}

\bibliographystyle{spmpsci}      
\bibliography{bibfile}   

\begin{thebibliography}{10}
\providecommand{\url}[1]{{#1}}
\providecommand{\urlprefix}{URL }
\expandafter\ifx\csname urlstyle\endcsname\relax
  \providecommand{\doi}[1]{DOI~\discretionary{}{}{}#1}\else
  \providecommand{\doi}{DOI~\discretionary{}{}{}\begingroup
  \urlstyle{rm}\Url}\fi

\bibitem{BroCoh17}
Brodutch, A., Cohen, E.: A scheme for performing strong and weak sequential
  measurements of non-commuting observables.
\newblock Quantum Studies: Mathematics and Foundations \textbf{4}(1), 13--27
  (2017).
\newblock \doi{10.1007/s40509-016-0084-8}.
\newblock \urlprefix\url{https://doi.org/10.1007/s40509-016-0084-8}

\bibitem{CorCuad17}
Cormann, M., Caudano, Y.: Geometric description of modular and weak values in
  discrete quantum systems using the majorana representation.
\newblock Journal of Physics A: Mathematical and Theoretical \textbf{50}(30),
  305,302 (2017).
\newblock \urlprefix\url{http://stacks.iop.org/1751-8121/50/i=30/a=305302}

\bibitem{CorEtAl16}
Cormann, M., Remy, M., Kolaric, B., Caudano, Y.: Revealing geometric phases in
  modular and weak values with a quantum eraser.
\newblock Phys. Rev. A \textbf{93}, 042,124 (2016).
\newblock \doi{10.1103/PhysRevA.93.042124}.
\newblock \urlprefix\url{https://link.aps.org/doi/10.1103/PhysRevA.93.042124}

\bibitem{GKP01}
Gottesman, D., Kitaev, A., Preskill, J.: Encoding a qubit in an oscillator.
\newblock Phys. Rev. A \textbf{64}, 012,310 (2001).
\newblock \doi{10.1103/PhysRevA.64.012310}.
\newblock \urlprefix\url{https://link.aps.org/doi/10.1103/PhysRevA.64.012310}

\bibitem{HoImoto16a}
Ho, L.B., Imoto, N.: Full characterization of modular values for
  finite-dimensional systems.
\newblock Physics Letters A \textbf{380}(25), 2129 -- 2135 (2016).
\newblock \doi{https://doi.org/10.1016/j.physleta.2016.05.005}.
\newblock
  \urlprefix\url{http://www.sciencedirect.com/science/article/pii/S0375960116301773}

\bibitem{HoImoto16b}
Ho, L.B., Imoto, N.: An interpretation and understanding of complex modular
  values.
\newblock arXiv:1602.01592 [quant-ph]  (2016).
\newblock \urlprefix\url{https://arxiv.org/abs/1602.01594}

\bibitem{HoImoto17}
Ho, L.B., Imoto, N.: Generalized modular-value-based scheme and its generalized
  modular value.
\newblock Phys. Rev. A \textbf{95}, 032,135 (2017).
\newblock \doi{10.1103/PhysRevA.95.032135}.
\newblock \urlprefix\url{https://link.aps.org/doi/10.1103/PhysRevA.95.032135}

\bibitem{KedemVaidman10}
Kedem, Y., Vaidman, L.: Modular values and weak values of quantum observables.
\newblock Phys. Rev. Lett. \textbf{105}, 230,401 (2010).
\newblock \doi{10.1103/PhysRevLett.105.230401}.
\newblock
  \urlprefix\url{https://link.aps.org/doi/10.1103/PhysRevLett.105.230401}

\bibitem{ParksSpence16}
Parks, A., Spence, S.: A pointer theory explanation of weak value persistence
  occurring in the quantum three box experimental data.
\newblock Acta Physica Polonica A \textbf{130}, 1265--1268 (2016)

\bibitem{ParksEtAl14}
Parks, A.D., Spence, S.E., Gray, J.E.: Exact pointer theories for von neumann
  projector measurements of pre- and postselected and preselected-only quantum
  systems: statistical mixtures and weak value persistence.
\newblock Proceedings of the Royal Society of London A: Mathematical, Physical
  and Engineering Sciences \textbf{470}(2162) (2014).
\newblock \doi{10.1098/rspa.2013.0651}.
\newblock
  \urlprefix\url{http://rspa.royalsocietypublishing.org/content/470/2162/20130651}

\bibitem{SpenceParks12}
Spence, S.E., Parks, A.D.: Experimental evidence for a dynamical non-locality
  induced effect in quantum interference using weak values.
\newblock Foundations of Physics \textbf{42}(6), 803--815 (2012).
\newblock \doi{10.1007/s10701-011-9596-6}.
\newblock \urlprefix\url{https://doi.org/10.1007/s10701-011-9596-6}

\bibitem{SpenceParks17}
Spence, S.E., Parks, A.D.: Experimental evidence for retro-causation in quantum
  mechanics using weak values.
\newblock Quantum Studies: Mathematics and Foundations \textbf{4}(1), 1--6
  (2017).
\newblock \doi{10.1007/s40509-016-0082-x}.
\newblock \urlprefix\url{https://doi.org/10.1007/s40509-016-0082-x}

\bibitem{SPN12}
Spence, S.E., Parks, A.D., Niemi, D.A.: Methods used to observe a dynamical
  quantum nonlocality effect in a twin mach-zehnder interferometer.
\newblock Appl. Opt. \textbf{51}(32), 7853--7857 (2012).
\newblock \doi{10.1364/AO.51.007853}.
\newblock \urlprefix\url{http://ao.osa.org/abstract.cfm?URI=ao-51-32-7853}

\end{thebibliography}

\end{document}